\documentclass[runningheads]{llncs}
\usepackage[T1]{fontenc}
\usepackage[numbers]{natbib}
\usepackage{graphicx}
\usepackage{hyperref}
\usepackage{subcaption}
\usepackage{colortbl, booktabs}
\usepackage{makecell}
\usepackage{wrapfig}
\usepackage{lineno}
\usepackage{framed}
\usepackage{mwe}
\usepackage{xcolor}

\usepackage{floatrow}
\newsavebox\CBox
\def\textBF#1{\sbox\CBox{#1}\resizebox{\wd\CBox}{\ht\CBox}{\textbf{#1}}}

\usepackage{floatrow}
\newfloatcommand{capbtabbox}{table}[][\FBwidth]

\usepackage{xcolor}
\definecolor{chelsea_colour}{HTML}{c260f0} 

\definecolor{tal_color}{HTML}{4d88ff}

\usepackage{algorithm}
\usepackage{algorithmicx}
\usepackage{algpseudocode}

\newcommand{\mat}[1]{\mathbf{\uppercase{#1}}}

\usepackage{amsmath}

\usepackage{bm}
\def\1{\bm{1}}

\def\vmu{{\bm{\mu}}}

\usepackage{array}
\newcolumntype{M}[1]{>{\centering\arraybackslash}m{#1}}
\usepackage{colortbl, booktabs}
\definecolor{Gray}{gray}{0.9}
\usepackage{multirow}

\usepackage{booktabs,arydshln}

\makeatletter
\def\adl@drawiv#1#2#3{%
        \hskip.5\tabcolsep
        \xleaders#3{#2.5\@tempdimb #1{1}#2.5\@tempdimb}%
                #2\z@ plus1fil minus1fil\relax
        \hskip.5\tabcolsep}
\newcommand{\cdashlinelr}[1]{%
  \noalign{\vskip\aboverulesep
           \global\let\@dashdrawstore\adl@draw
           \global\let\adl@draw\adl@drawiv}
  \cdashline{#1}
  \noalign{\global\let\adl@draw\@dashdrawstore
           \vskip\belowrulesep}}
\makeatother

\begin{document}
\title{Heatmap Regression for Lesion Detection using Pointwise Annotations}

\author{Chelsea Myers-Colet\inst{1}\thanks{Corresponding author} \and
Julien Schroeter\inst{1} \and
Douglas L. Arnold\inst{2} \and
Tal Arbel\inst{1}}
%

%index{Myers-Colet, Chelsea}
%index{Schroeter, Julien}
%index{Arnold, Douglas L.}
%index{Arbel, Tal}

\authorrunning{C. Myers-Colet et al.}
\institute{Centre for Intelligent Machines, McGill University, Montreal, Canada \\
\email{\{cmyers, julien, arbel\}@cim.mcgill.ca}\\
\and
Montreal Neurological Institute, McGill University, Montreal, Canada\\
\email{douglas.arnold@mcgill.ca}}

\maketitle              % typeset the header of the contribution
\begin{abstract}
In many clinical contexts, detecting all lesions is imperative for evaluating disease activity. Standard approaches pose lesion detection as a segmentation problem despite the time-consuming nature of acquiring segmentation labels. In this paper, we present a lesion detection method which relies only on point labels. Our model, which is trained via heatmap regression, can detect a variable number of lesions in a probabilistic manner. In fact, our proposed post-processing method offers a reliable way of directly estimating the lesion existence uncertainty. Experimental results on Gad lesion detection show our point-based method performs competitively compared to training on expensive segmentation labels. Finally, our detection model provides a suitable pre-training for segmentation. When fine-tuning on only~17 segmentation samples, we achieve comparable performance to training with the full dataset.

\end{abstract}

\section{Introduction}
For many diseases, detecting the presence and location of all lesions is vital for estimating disease burden and treatment efficacy. In stroke patients, for example, the location of a cerebral hemorrhage was shown to be an important factor in assessing the risk of aspiration~\cite{daniels1999lesion} thus, failing to locate even a single one could drastically impact the assessment.
Similarly, in patients with Multiple Sclerosis~(MS), detecting and tracking all gadolinium-enhancing lesions~(Gad lesions), whether large or small, is especially relevant for determining treatment response in clinical trials~\cite{rudick2004defining}. Detecting all Gad lesions is imperative as just one new lesion indicates new disease activity.

To achieve this goal, standard practice in deep learning consists of training a lesion segmentation model with a post-processing detection step~\cite{lundervold2019overview,doyle2017lesion}. However, segmentation labels are expensive and time consuming to acquire. To this end, we develop a lesion detection model trained on pointwise labels thereby reducing the manual annotation burden. Unlike previous point annotation-based methods~\cite{sharan2021point,stern2021heatmap,thaler2021modeling},
ours combines the ability to detect a variable number of lesions with the benefit of leveraging a probabilistic approach. Indeed, our refinement method is not only independent of a specific binarization threshold, it offers a unique way of estimating the lesion existence probability.
Our contributions are threefold:
\begin{enumerate}
\item[\textbf{(1)}] We demonstrate the merit of training on point annotations via heatmap regression over segmentation labels on the task of Gad lesion detection. With weaker labels, our models still achieve better detection performance. 
\item[\textbf{(2)}] Our proposed refinement method allows for a reliable estimation of lesion existence uncertainty thus providing valuable feedback for clinical review.
\item[\textbf{(3)}] When the end goal is segmentation, our detection models provide a suitable pre-training for fine-tuning on a limited set of segmentation labels. When having access to only 17 segmentation samples, we can achieve comparable performance to a model trained on the entire segmentation dataset.
\end{enumerate}

\section{Related Work}
Point annotations are often extremely sparse which leads to instability during training of deep neural networks. Therefore, most state-of-the-art methods rely on the application of a smoothing operation to point labels. A Gaussian filter is commonly applied to create a heatmap as was done in~\cite{stern2021heatmap, sharan2021point} for suture detection.
Others have found success applying distance map transformations. For instance, \citet{han2021detecting} and \citet{van2019automated} used Euclidean and Geodesic distance maps to perform lesion detection.
We demonstrate the benefits of training with Gaussian heatmaps over distance maps as they offer a more precise and interpretable probabilistic prediction yielding superior detection performance.

Irrespective of the choice of smoothing used for training, detection methods will often differ in their post-processing refinement step, i.e. in extracting lesion coordinates from a 
predicted 
heatmap. The simplest approach consists in finding the location with the maximum mass~\cite{donne2016mate,chen2018ccdn,sharan2021point} or computing the centre of mass~\cite{stern2021heatmap}. Although these approaches easily allow for the detection of multiple lesions, they require careful tuning of the binarization threshold and are susceptible to missing both isolated and overlapping peaks. More sophisticated methods exist which aim to fit a Gaussian distribution to the predicted heatmap thus retaining its probabilistic interpretation, e.g.~\cite{zhang2020distribution,graving2019deepposekit}. Specifically, to perform cephalogram landmark detection \citet{thaler2021modeling} align a Gaussian distribution via Least Squares curve fitting.
Since the approach taken in~\cite{thaler2021modeling} is limited to a set number of landmarks, we extend it to detect a variable number of lesions. Our method thus offers the flexibility of simpler approaches, without any dependence on a binarization threshold, while providing a probabilistic interpretation.

\section{Method}

In this work, we propose a strategy to detect the presence and location of multiple lesions from brain MRIs of patients with a neurodegenerative disease. Our model is trained via heatmap regression~(Section~\ref{sec:train}) while lesion detection is performed in a post-processing step~(Section~\ref{sec:detect}). Finally, we present a transfer learning scheme to perform segmentation on a limited dataset~(Section~\ref{sec:method-transfer}).

\subsection{Training via Heatmap Regression}
\label{sec:train}
The proposed heatmap regression training scheme requires a domain expert to label only a single point identifying each lesion, e.g. by marking the approximate centre of the lesion. To stabilize training, we apply a Gaussian filter with smoothing parameter~$\sigma$ to the point annotations thus creating a multi-instance heatmap~\cite{stern2021heatmap,thaler2021modeling,sharan2021point,wang2019adaptive,pfister2015flowing,hervella2020deep}. 
Since all lesions are represented by a single point and smoothed using the same value of~$\sigma$, equal importance is attributed to lesions of all sizes. 
We train a model $f_{\theta}$ to map a sequence of input MRIs to a predicted heatmap $\hat{\mat{H}}_i$.

\subsection{Detection During Inference}
\label{sec:detect}

\begin{figure}[t]
\begin{floatrow}
     \centering
     \includegraphics[width=1.0\textwidth]{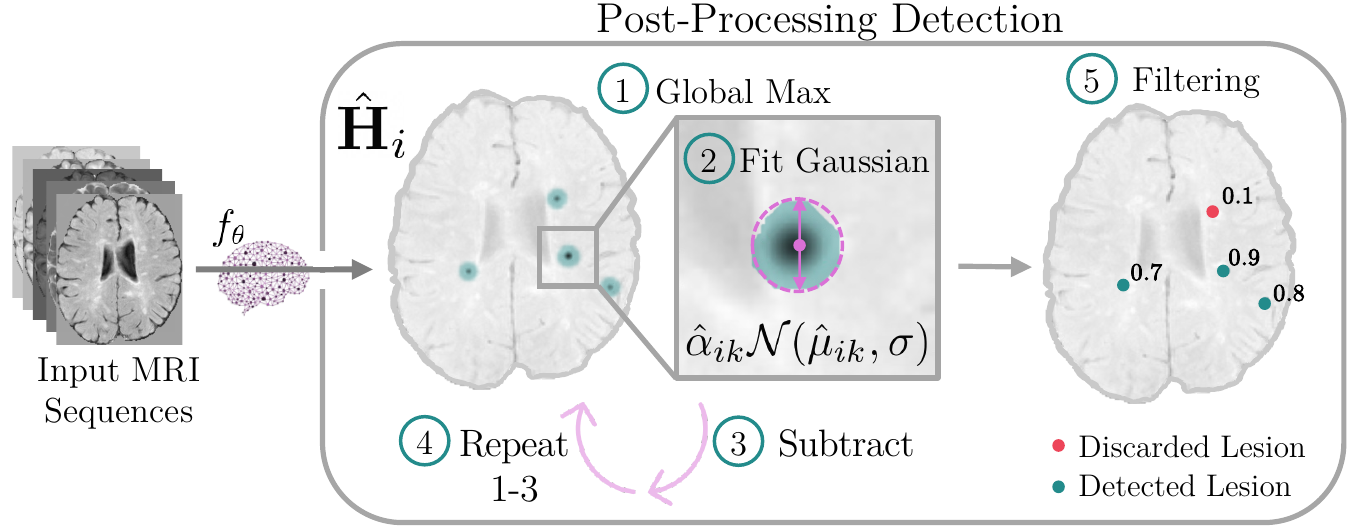}
        \caption{Overview of detection method given a predicted heatmap~$\hat{\mat{H}}_i$:
        lesion candidates are found by (1) locating the global maximum, (2) fitting a Gaussian distribution to an extracted region and (3) subtracting the influence of this lesion from the
        heatmap. (4) Repeat steps 1-3 before (5) filtering out unlikely lesions.
        }
    \label{fig:main}
\end{floatrow}
\end{figure}

Given a continuous heatmap~$\hat{\mat{H}}_i$, we now aim to detect individual lesions. Specifically, for patient~$i$, we wish to represent the $k^{th}$ detected lesion by a single point,~$\hat{\vmu}_{ik}$, which can be extracted from the heatmap. We assume the predicted heatmap~$\hat{\mat{H}}_i$ will model a sum of Gaussian distributions~(each describing a single lesion) to mimic the target heatmap:\footnote{Valid as long as~$f_{\theta}$ sufficiently minimizes the loss and thus models the target.}
\begin{equation}
    \begin{split}
        \hat{\mat{H}}_i = \sum_K \hat{\mat{H}}_{ik} = \sum_K \mathcal{N}(
    \hat{\vmu}_{ik}, \sigma)
    \end{split}
    \label{eq:sum}
\end{equation}
Our method essentially aims to find the individual Gaussian distributions comprising the sum in equation~\ref{eq:sum} in an iterative manner as shown in Figure~\ref{fig:main}. We now describe each depicted step in detail.

\paragraph{\textbf{(1) Locate Global Maximum}}
The location of the global maximum serves as an initial estimate for the~$k^{th}$ predicted lesion centre,~$\hat{\vmu}_{ik}$.

\paragraph{\textbf{(2) Gaussian Fitting}}
In the region surrounding a detected lesion with centre~$\hat{\vmu}_{ik}$, we fit a Gaussian distribution with normalizing constant $\hat{\alpha}_{ik}$: 
\begin{equation}
    \begin{split}
        \hat{\mat{H}}_{ik} = \hat{\alpha}_{ik} \mathcal{N}(\hat{\vmu}_{ik}, \sigma)
    \end{split}
\end{equation}
Provided there is minimal overlap between neighbouring lesions, we can use a Least Squares curve fitting algorithm to estimate~$\hat{\alpha}_{ik}$ and $\hat{\vmu}_{ik}$.\footnote{ 
Valid for Gad lesions given a sufficiently small smoothing parameter~$\sigma$.}
The normalizing constant,~$\hat{\alpha}_{ik}$, represents the prior probability of producing a peak in this region~(from Bayes' Theorem), i.e. it is the belief that a lesion exists in the given region. We thus refer to~$\hat{\alpha}_{ik}$ as the lesion existence probability~(similar to~\cite{ourpaper2022}). As an initial estimate for $\hat{\alpha}_{ik}$, we sum within the extracted region, i.e. the hypothesis space, 
as shown in Figure~\ref{fig:main} (2). 

\paragraph{\textbf{(3) Subtract}}
Now that potential lesion~$k$ has been identified and fitted with a continuous Gaussian function, we remove its contribution to the sum in equation~\ref{eq:sum}. This allows our method to more easily detect the individual contributions of neighbouring lesions with overlapping Gaussian distributions.

\begin{equation}
    \begin{split}
        \hat{\mat{H}}^\prime_i = \hat{\mat{H}}_i - \hat{\alpha}_{ik} \mathcal{N}(\hat{\vmu}_{ik}, \sigma)
    \end{split}
\end{equation}

\paragraph{\textbf{(4) Repeat}}
Since we have subtracted the contribution of lesion~$k$ from the aggregated heatmap, the global maximum now corresponds to a different candidate lesion. Steps 1 to 3 are repeated until a maximum number of lesions have been found or when the lesion existence probability drops below a threshold, e.g.~0.01.

\paragraph{\textbf{(5) Filtering}}
Lesions with a low probability of existence are discarded~(threshold optimized on the validation set). By overestimating the lesion count and subsequently discarding regions unlikely to contain a lesion, we can better account for noisy peaks in the heatmap. We 
evaluate the calibration of these probabilities and demonstrate the validity of this filtering step~(Section~\ref{sec::lesion-existence-calibration}).

\subsection{Segmentation Transfer Learning}
\label{sec:method-transfer}
In addition to detection, estimating a lesion segmentation can be beneficial for assessing lesion load.
We therefore design a transfer learning scheme which first relies on building a strong lesion detector using point annotations before fine-tuning on a small segmentation dataset. Specifically, we (1) train a detection model on point annotations until convergence; (2) build a small segmentation training set; (3) fine-tune the detection model on segmentation samples only. Training in this manner minimizes the amount of detailed segmentation labels that must be generated.

\section{Experiments and Results}
The proposed heatmap regression model is compared against three benchmarks in terms of detection performance.
We train models on (1) segmentation labels, (2) Euclidean distance maps and (3) Geodesic distance maps~\cite{han2021detecting,van2019automated}. 
Similar to our method, lesions are detected from the output prediction in a post-processing step. Here, we instead binarize the output at threshold~$\tau$~(optimized on the validation set), cluster connected components to form detected lesions and use the centre of mass~(segmentation) or the maximum~(detection) to represent the lesion~(referred to as {\sc CC}). As an additional benchmark, we apply this method to heatmap outputs from our proposed regression models. 
This is in line with detection methods used by~\cite{nair2020exploring,de2018automated} for segmentation outputs and~\cite{sharan2021point,stern2021heatmap} for heatmap predictions.

\subsection{Experimental Setup}
\subsubsection{Dataset} We evaluate our method on Gad lesion detection as they are a relevant indicator of disease activity in MS patients~\cite{mcfarland1992using}. However, their subtlety and extreme size variation makes them difficult to identify. Experiments are performed using a large, multi-centre, multi-scanner proprietary dataset consisting of 1067 patients involved in a clinical trial to treat Relapsing-Remitting MS. Multi-modal MRIs, including post-contrast T1-weighted MRI, are available for each patient and are provided as inputs to our system.
For fairness, we create train~(60\%), validation~(20\%) and test~(20\%) sets by first splitting at the patient level. 
We have access to manually derived Gad lesion segmentation masks. Each sample is first independently rated by two experts who then meet to produce a consensus. 
Point labels were generated directly from segmentation masks by calculating the centre of mass of each lesion and transformed into either heatmaps, using a Gaussian kernel with smoothing parameter~$\sigma$, or distance maps~(baseline methods), using decay parameter~$p$. Hyperparameters were selected based on validation performance. 

\subsubsection{Model} We train a modified 5-layer U-Net~\cite{ronneberger2015u} with dropout and instance normalization using a Mean-Squared Error loss for heatmap regression and a weighted cross-entropy loss for segmentation. See code for details\footnote{\url{https://github.com/ChelseaM-C/MICCAI2022-Heatmap-Lesion-Detection}}.

\subsubsection{Evaluation} 
We apply the Hungarian algorithm~\cite{kuhn1955hungarian} to match predicted lesions to ground truth lesions using Euclidean distance as a cost metric. Assignments with large distances are considered both a false positive and a false negative. 

\subsection{Lesion Detection Results}
Despite only having access to point annotations, the proposed Gaussian heatmap approach performs competitively with the segmentation 
baseline~(see Table~\ref{tbl::results::test-detection}). In fact, our proposed iterative detection method~({\sc Gaussian}) even slightly outperforms the segmentation model on all detection metrics. By contrast, both distance map approaches show notably worse performance with especially low recall scores indicating a high number of missed lesions. 
The proposed model additionally outperforms competing methods for the task of small lesion detection~(3 to 10 voxels in size) underlining the merit of training directly for detection. Segmentation models will typically place more importance on larger lesions since they have a higher contribution to the loss, a bias not imposed by our detection model. Our model additionally does not sacrifice precision for high recall on small lesions; we perform on par with segmentation.

\begin{table}[t]
\caption{Lesion detection results as a mean over 3 runs. Reported is the detection {\sc F1}-score, precision, recall and small lesion recall for models trained with segmentation, Gaussian heatmap or distance map~(Geodesic, Euclidean)~\cite{han2021detecting,van2019automated} labels using connected components~({\sc CC}) or Gaussian fitting~({\sc Gaussian}).} 
\label{tbl::results::test-detection}
\setlength{\tabcolsep}{1pt}
%\label{tbl::results::test-detection}
\begin{center}
\begin{small}
\begin{sc}
\resizebox{0.99\textwidth}{!}{

\begin{tabular}{ll}
%\white{.}\\
\toprule
%\multicolumn{2}{l}
% \multicolumn{2}{l}{Label Type} %& $\lambda_{sum}$ 
Label Type &\makecell{\phantom{.} \\ \phantom{.}}
\\
\midrule
Segmentation & \\ 
\cdashlinelr{1-2}
Euclidean Map &  \\ 
\cdashlinelr{1-2}
Geodesic Map &  \\ 
\cdashlinelr{1-2}
%\makecell[l]{Gaussian Heatmap \\ ~~~($\sigma=1.0$)} 
\multirow{2}{*}{Gaussian Heatmap}  &                  \\
& \\

\bottomrule
\end{tabular}

\setlength{\tabcolsep}{3pt}
\rule{-0.8em}{0ex}

\begin{tabular}{ccccc}
%\white{.}\\
\toprule
%\multicolumn{2}{l}
\makecell{Detection \\ Method} & \makecell{F1- \\ score} & \makecell{Precision} & \makecell{Recall} & \makecell{Small Lesion \\ Recall} \\
\midrule
% CC & 85.4 & 85.3 & 85.5 & 67.7 \\ 
% \cdashlinelr{1-5}
% CC & 80.6 & \textbf{92.6} &  71.4 & 51.0 \\ 
% \cdashlinelr{1-5}
% CC & 73.7 & 81.0 & 67.8 & 47.8 \\ 
% \cdashlinelr{1-5}
% CC & 83.9 & 80.9 & 87.3 & \textbf{75.0} \\ 
% Gaussian & \textbf{86.3} & 87.0 & \textbf{85.7} & 70.4 \\

CC & 85.4 $\pm$ 0.02 & 85.3 $\pm$ 1.10 & 85.5 $\pm$ 1.15 & 67.7 $\pm$ 3.58 \\ 
\cdashlinelr{1-5}
CC & 80.6 $\pm$ 1.01 & \textbf{92.6} $\pm$ \textbf{1.28}  &  71.4 $\pm$ 2.14 & 51.0 $\pm$ 3.28\\ 
\cdashlinelr{1-5}
CC & 73.7 $\pm$ 4.89 & 81.0 $\pm$ 7.97 & 67.8 $\pm$ 2.98 & 47.8 $\pm$ 2.81\\ 
\cdashlinelr{1-5}
CC & 83.9 $\pm$ 0.27 & 80.9 $\pm$ 4.43 & 87.3 $\pm$ 2.64 & \textbf{75.0} $\pm$ \textbf{5.75} \\ 
Gaussian & \textbf{86.3} $\pm$ \textbf{0.24} & 87.0 $\pm$ 1.89 & \textbf{85.7} $\pm$ \textbf{1.44} & 70.4 $\pm$ 4.47\\

\bottomrule
\end{tabular}
}	
\end{sc}
\end{small}
\end{center}
\vspace{-0.4cm}

\end{table}

\subsubsection{Lesion Existence Probability Evaluation} 
We evaluate the quality of our fitted lesion existence probabilities on the basis of calibration and derived uncertainty to justify both the curve fitting and filtering steps.

\begin{figure}[t]
\centering
 \begin{subfigure}[t]{0.48\textwidth}
    \begin{center}
        \includegraphics[width=\linewidth]{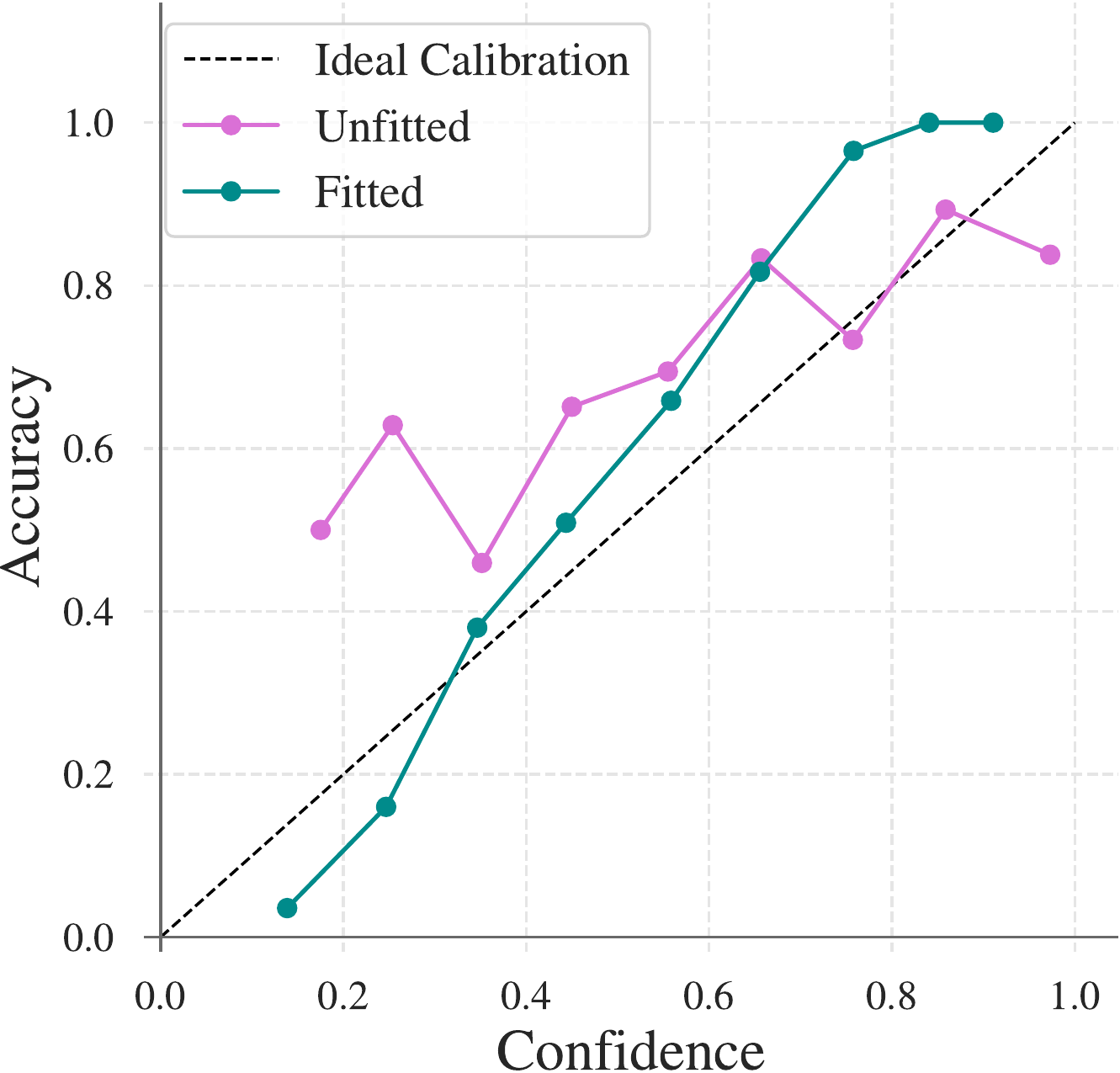}
        \caption{
        }%
        \label{fig:calibration}
    \end{center}
 \end{subfigure}
  \hfill
  \begin{subfigure}[t]{0.50\textwidth}
  \centering
      \includegraphics[width=\linewidth]{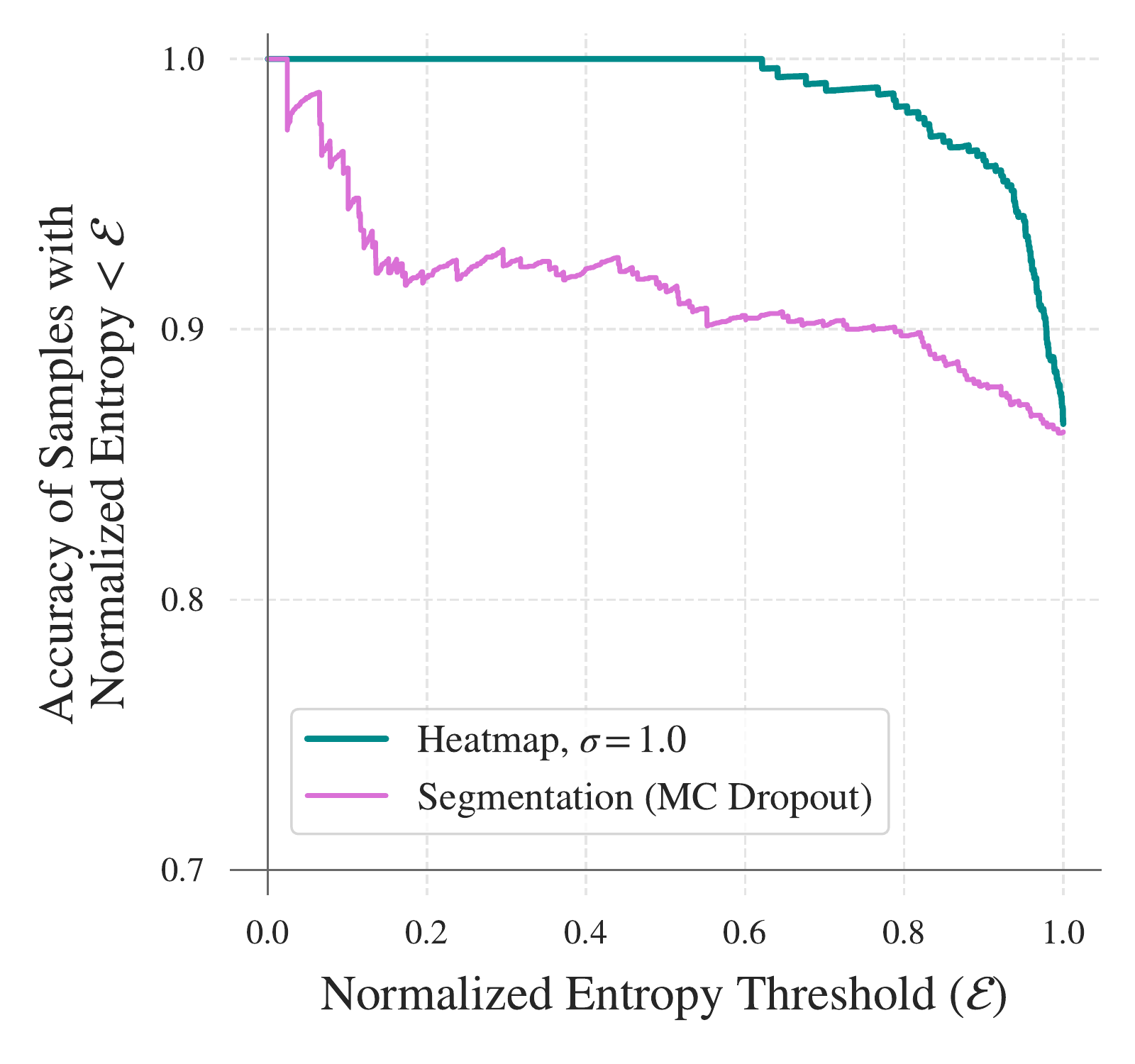}
      \caption{
      }
      \label{fig:uncertainty-lt}
 \end{subfigure}
  \caption{Lesion existence probability evaluation. (a) Calibration of unfitted~(pink) vs. fitted~(green)
  probabilities.
  (b) Detection accuracy of least uncertain samples according to our model~(green) vs. 
      MC Dropout applied to segmentation~(pink).
}
\label{fig:uncertainty}
\end{figure}

\paragraph{\textbf{(1) Calibration}}
\label{sec::lesion-existence-calibration}
We compare the calibration~\cite{guo2017calibration} of the lesion existence probabilities before and after Least Squares curve fitting. Recall the initial estimate for~$\alpha_{ik}$ is found by summing locally within the extracted region.  
Our proposed existence probabilities are well calibrated~(Figure~\ref{fig:calibration}), with little deviation from the ideal case thus justifying our proposed filtering step. As well, the fitted probabilities are significantly better calibrated than the initial estimates demonstrating the benefit of curve fitting.

\paragraph{\textbf{(2) Uncertainty}} 
While it is important to produce accurate predictions, quantifying their uncertainty is of equal importance in the medical domain. 
We can compute the entropy of our lesion existence probabilities without sampling and show it is well correlated with detection accuracy.
As we consider only the least uncertain instances, we observe a monotonic increasing trend, even achieving an accuracy of~100\%~(Figure~\ref{fig:uncertainty-lt}). 
Similar results are achieved with the more standard MC Dropout approach~\cite{gal2016dropout} applied to segmentation outputs~(calculated at a lesion level as in~\cite{nair2020exploring}). 

Our derived lesion existence probabilities are not only well-calibrated, they produce meaningful uncertainty estimates. With only a single forward pass, our uncertainty estimates perform on par with standard sampling-based approaches. 

\subsection{Lesion Segmentation via Transfer Learning}
To demonstrate the adaptability of our method, we fine-tune the trained heatmap regression models with a small segmentation dataset as described in Section~\ref{sec:method-transfer}. Specifically, we use segmentation labels for a randomly chosen 1\% of our total training set for fine-tuning. 
To account for bias in the selected subset, we repeat this process~3 times and average the results.
For comparison, we train from scratch with this limited set as well as on the full segmentation training set. We additionally include results on random subsets of~5\% and~10\% in the appendix along with the associated standard deviation of each experiment.

\begin{table}
\caption{Segmentation transfer learning results averaged over 3 random subsets. We present segmentation F1-score~({\sc Seg F1.}) and detection~({\sc Det.}) metrics on the fine-tuned segmentation models: F1-score, precision, recall.
Pre-trained models are distinguished by their smoothing hyperparameter~$\sigma$. 
} 
\label{tbl::results::test-transfer}
\setlength{\tabcolsep}{3pt}
\begin{center}
\begin{small}

\resizebox{\textwidth}{!}{
\begin{tabular}{ll}
%\white{.}\\
\toprule
\sc{Quantity} & \sc{Pre-Trained} \\
\sc{Seg. Labels} & \sc{Model} \\
\midrule
\textbf{100\%} & {\footnotesize None} \\
\midrule
\multirow{4}{*}{\textbf{1\%}} & {\footnotesize None} \\
                                        & \textBF{$\sigma=1.0$} \\
                                        & \textBF{$\sigma=1.25$} \\
                                       & \textBF{$\sigma=1.5$} \\
                            & {\footnotesize \sc{Euclidean}} \\
                            & {\footnotesize \sc{Geodesic}} \\
                            
\bottomrule
\end{tabular}

\setlength{\tabcolsep}{5pt}
\rule{-0.8em}{0ex}
\begin{sc}
\begin{tabular}{cccc}

\toprule
Seg. F1 & Det. F1 & Det. & Det. \\%& Small Lesion \\
 & & Precision & Recall\\ %& Recall \\
\midrule
70.5 $\pm$ 0.31 & 85.4 $\pm$ 0.02 & 85.3 $\pm$ 1.10 & 85.5 $\pm$ 1.15 \\%& 64.3 \\
\midrule
60.7 $\pm$ 2.48 & 69.7 $\pm$ 5.92 & 79.0 $\pm$ 6.47 & 62.7 $\pm$ 7.69 \\%& 36.9   \\
67.2 $\pm$ 0.71 & \textbf{85.4} $\pm$ \textbf{0.62} & 83.9 $\pm$ 1.74 & \textbf{87.0}  $\pm$ \textbf{2.40}\\%& 69.0\\ 
\textbf{67.6} $\pm$ \textbf{1.31} & 84.5 $\pm$ 0.38  & \textbf{86.0} $\pm$ \textbf{0.67} & 83.1 $\pm$ 1.32\\%& \textbf{71.4} \\
67.2 $\pm$ 1.51 & 85.0 $\pm$ 0.50 & 83.6 $\pm$ 1.01 & 86.6 $\pm$ 1.10 \\%& \textbf{75.6}\\
66.3 $\pm$ 0.18 & 84.7 $\pm$ 1.45  & 83.3 $\pm$ 3.24 & 86.1 $\pm$ 1.67 \\
59.9 $\pm$ 1.44 & 77.9 $\pm$ 4.36  & 85.6 $\pm$ 7.33 & 71.7 $\pm$ 3.49 \\

\bottomrule

\end{tabular} 
\end{sc}
}

\end{small}
\end{center}

\end{table}

Remarkably, our pre-trained models show only a~3\% drop in segmentation F1-score performance with a mere~1\% of the segmentation labels compared to the model trained on the full segmentation dataset~(see Table~\ref{tbl::results::test-transfer}).
By contrast, the model trained from scratch with the same~1\% of segmentation labels shows a~10\% drop in segmentation F1-score. This emphasizes the importance of detecting lesions since models trained from scratch in the low data regime show considerably lower detection F1-score. It is clear the models do not require very much data in order to properly segment lesions as demonstrated by competitive performance of our pre-trained models. However, as indicated by poor detection performance of the pure segmentation models in the low data regime, 
it is clear these models need help \text{localizing} lesions before they can be segmented. We can see a similar trend with the models pre-trained on distance maps. The Euclidean distance maps offered higher detection scores than the Geodesic ones~(Table~\ref{tbl::results::test-detection}) and therefore serve as a better pre-training for segmentation, although still lower than our models.

\section{Discussion and Conclusion}
In this work, we have demonstrated how training a heatmap regression model to detect lesions
can achieve the same, and at times better, detection performance compared to a segmentation model.
By requiring clinicians to indicate a single point within each lesion,
% By simply requiring clinicians to indicate approximate lesion centres, 
our approach significantly reduces the annotation burden imposed by deep learning segmentation methods.
Our proposed 
method of iteratively fitting Gaussian distributions to a predicted heatmap produces well-calibrated existence probabilities which capture the underlying uncertainty. 

Perhaps most significantly, our transfer learning experiments have revealed an important aspect about segmentation models. 
Our results demonstrate that segmentation models must learn first and foremost to find lesions.
Indeed, our models, which are already adept at lesion detection, can easily learn to delineate borders with only a few segmentation samples. By contrast, the models provided with the same limited set of segmentation labels trained from scratch fail primarily to detect lesions thus lowering their segmentation scores. 
It therefore presents an unnecessary burden on clinicians to require them to
manually segment large datasets in order to build an accurate deep learning segmentation model.

Although we have demonstrated many benefits, Gaussian heatmap matching has its limitations.
The smoothing hyperparameter requires careful tuning to both maintain stable training and to avoid a significant overlap in peaks~(especially for densely packed lesions).
As well, the method still requires an expert annotator to mark approximate lesion centres however, this is much less time-consuming than fully outlining each lesion. We recognize this could introduce high variability in the labels regarding where the point is placed within each lesion. Though the current model was trained on precise centres of mass, the proposed method does not necessarily impose any such constraints, in theory. Future work is needed to evaluate the robustness of the model to high variability in the label space.

In summary, our proposed training scheme and iterative Gaussian fitting post-processing step constitute an accurate and label-efficient method of performing lesion detection and segmentation.

\subsubsection{Acknowledgements}

This work was supported by awards from the International Progressive MS Alliance (PA-1412-02420), the Canada Institute for Advanced Research (CIFAR) Artificial Intelligence Chairs program (Arbel), the Canadian Natural Science and Engineering Research Council (CGSM-NSERC-2021-Myers-Colet) and the Fonds de recherche du Québec (303237). The authors would  also like to thank Justin Szeto, Kirill Vasilevski, Brennan Nichyporuk and Eric Zimmermann as well as the companies who provided the clinical trial data: Biogen, BioMS, MedDay, Novartis, Roche/Genentech, and Teva. Supplementary computational resources were provided by Calcul Québec, WestGrid, and Compute Canada.

\bibliography{paper1682}

\clearpage

\section*{Appendix}

\label{appendix}

\begin{table}
\caption{Lesion detection results by size with standard deviation. Reported is the detection recall for small, medium and large lesions for models trained with segmentation, Gaussian heatmap or distance map~(Geodesic, Euclidean) labels using connected components~({\sc CC}) or Gaussian fitting~({\sc Gaussian}).} 
\label{app:tbl:detection-size}
\setlength{\tabcolsep}{5pt}
%\label{app:tbl:deteciton-size}
\begin{center}
\begin{small}
\begin{sc}
\resizebox{0.99\textwidth}{!}{
\begin{tabular}{ll}
%\white{.}\\
\toprule
%\multicolumn{2}{l}
% \multicolumn{2}{l}{Label Type} %& $\lambda_{sum}$ 
Label Type &\makecell{\phantom{.} \\ \phantom{.}}
\\
\midrule
Segmentation & \\ 
\cdashlinelr{1-2}
Euclidean Map &  \\ 
\cdashlinelr{1-2}
Geodesic Map &  \\ 
\cdashlinelr{1-2}
%\makecell[l]{Gaussian Heatmap \\ ~~~($\sigma=1.0$)} 
\multirow{2}{*}{Gaussian Heatmap}  &                  \\
& \\

\bottomrule
\end{tabular}

\setlength{\tabcolsep}{3pt}
\rule{-0.8em}{0ex}

\begin{tabular}{cccc}
%\white{.}\\
\toprule
%\multicolumn{2}{l}
\makecell{Detection \\ Method} & \makecell{Small Lesion \\ Recall } & \makecell{Medium Lesion\\ Recall}  & \makecell{Large Lesion \\ Recall} \\
\midrule
CC & 67.7 $\pm$ 3.58 & 89.6 $\pm$ 2.35 & 97.3 $\pm$ 3.76 \\ 
\cdashlinelr{1-4}
CC & 51.0 $\pm$ 3.28 & 77.1 $\pm$ 1.94 &  89.3 $\pm$ 2.80 \\ 
\cdashlinelr{1-4}
CC & 47.8 $\pm$ 2.81 & 73.9 $\pm$ 3.90 & 83.8 $\pm$ 4.01  \\ 
\cdashlinelr{1-4}
CC & \textbf{75.0} $\pm$ \textbf{5.75} & \textbf{91.6} $\pm$ \textbf{1.74} & 95.6 $\pm$ 1.11  \\ 
Gaussian & 70.4 $\pm$ 4.47 & 90.7 $\pm$ 0.63 & \textbf{97.8} $\pm$ \textbf{1.11}  \\
%  \cdashlinelr{3-6}                                                        

\bottomrule
\end{tabular}
}	
\end{sc}
\end{small}
\end{center}
%\vspace{-0.4cm}

\end{table}

\vspace{-0.5cm}

\begin{table}[H]
\caption{Segmentation transfer learning results with standard deviation averaged over 3 random subsets. We present segmentation F1-score~({\sc Seg F1.}) and detection~({\sc Det.}) metrics on the fine-tuned segmentation models: F1-score, precision, recall.
Pre-trained models are distinguished by their smoothing hyperparameter~$\sigma$. 
} 
\label{tbl::results::test-transfer3}
\setlength{\tabcolsep}{3pt}
\begin{center}
\begin{small}

\resizebox{1.0\textwidth}{!}{
\begin{tabular}{ll}
%\white{.}\\
\toprule
\sc{Quantity} & \sc{Pre-Trained} \\
\sc{Seg. Labels} & \sc{Model} \\
\midrule
\textbf{100\%} & {\footnotesize None} \\
\midrule
\multirow{4}{*}{\textbf{1\%}} & {\footnotesize None} \\
                                        & \textBF{$\sigma=1.0$} \\
                                        & \textBF{$\sigma=1.25$} \\
                                      & \textBF{$\sigma=1.5$} \\
                            & {\footnotesize \sc{Euclidean}} \\
                            & {\footnotesize \sc{Geodesic}} \\
\cdashlinelr{1-2} 
\multirow{4}{*}{\textbf{5\%}} & {\footnotesize None} \\
                                        & \textBF{$\sigma=1.0$} \\
                                        & \textBF{$\sigma=1.25$} \\
                                      & \textBF{$\sigma=1.5$} \\
                            & {\footnotesize \sc{Euclidean}} \\
                            & {\footnotesize \sc{Geodesic}} \\
\cdashlinelr{1-2} 
\multirow{4}{*}{\textbf{10\%}} & {\footnotesize None} \\
                                        & \textBF{$\sigma=1.0$} \\
                                        & \textBF{$\sigma=1.25$} \\
                                      & \textBF{$\sigma=1.5$} \\
                            & {\footnotesize \sc{Euclidean}} \\
                            & {\footnotesize \sc{Geodesic}} \\
\bottomrule
\end{tabular}

\setlength{\tabcolsep}{5pt}
\rule{-0.8em}{0ex}
\begin{sc}
\begin{tabular}{cccc}

%\multicolumn{2}{c}{\textBF{Jaccard}}\\
\toprule
Seg. F1 & Det. F1  & Det. & Det. \\%& Small Lesion \\
& & Precision & Recall\\ %& Recall \\
\midrule
70.5 $\pm$ 0.31 & 85.4 $\pm$ 0.02 & 85.3 $\pm$ 1.10 & 85.5 $\pm$ 1.15 \\%& 64.3 \\
\midrule
60.7 $\pm$ 2.48 & 69.7 $\pm$ 5.92 & 79.0 $\pm$ 6.47 & 62.7 $\pm$ 7.69 \\%& 36.9   \\
67.2 $\pm$ 0.71 & \textbf{85.4} $\pm$ \textbf{0.62} & 83.9 $\pm$ 1.74 & \textbf{87.0}  $\pm$ \textbf{2.40}\\%& 69.0\\ 
\textbf{67.6} $\pm$ \textbf{1.31} & 84.5 $\pm$ 0.38  & \textbf{86.0} $\pm$ \textbf{0.67} & 83.1 $\pm$ 1.32\\%& \textbf{71.4} \\
67.2 $\pm$ 1.51 & 85.0 $\pm$ 0.50 & 83.6 $\pm$ 1.01 & 86.6 $\pm$ 1.10 \\%& \textbf{75.6}\\
66.3 $\pm$ 0.18 & 84.7 $\pm$ 1.45  & 83.3 $\pm$ 3.24 & 86.1 $\pm$ 1.67 \\
59.9 $\pm$ 1.44 & 77.9 $\pm$ 4.36  & 85.6 $\pm$ 7.33 & 71.7 $\pm$ 3.49 \\
 \cdashlinelr{1-4} 
% \\
% \\
%  \cdashlinelr{1-9} 
67.1 $\pm$ 1.07 & 81.7 $\pm$ 0.80 & 84.5 $\pm$ 1.74 & 79.2 $\pm$ 2.08 \\%& 51.2  \\
68.9 $\pm$ 0.006 & \textbf{86.2} $\pm$ \textbf{0.002}  & \textbf{87.8} $\pm$ \textbf{0.16} & 84.6 $\pm$ 0.61 \\%& 66.7 \\ 
69.1 $\pm$ 0.31 & 84.9 $\pm$ 0.49 & 86.4 $\pm$ 1.83 & 83.5 $\pm$ 0.80 \\%& 66.1 \\
\textbf{69.5} $\pm$ \textbf{0.54} & 84.5 $\pm$ 0.73 & 82.7 $\pm$ 3.63 & \textbf{86.6} $\pm$ \textbf{2.59} \\%& \textbf{67.3} \\
68.4 $\pm$ 0.70 & 85.4 $\pm$ 0.49 & 84.8 $\pm$ 0.48 & 86.0 $\pm$ 0.70 \\
62.8 $\pm$ 1.68 & 83.2 $\pm$ 1.40 & 85.5 $\pm$ 2.08 & 81.0 $\pm$ 1.57 \\
 \cdashlinelr{1-4} 
% \\
% \\
%  \cdashlinelr{1-9} 
\textbf{70.2} $\pm$ \textbf{0.85} & 84.3 $\pm$ 0.77 & 83.9 $\pm$ 0.75 & 84.7 $\pm$ 1.09 \\%& 51.2  \\
69.3 $\pm$ 0.36 & 85.4 $\pm$ 0.53 & \textbf{85.8} $\pm$ \textbf{2.57} & 85.1 $\pm$ 2.69 \\%& 66.7 \\ 
69.2 $\pm$ 0.32 & 84.9 $\pm$ 0.99 & 84.9 $\pm$ 2.36 & 84.8 $\pm$ 0.68 \\%& 66.1 \\
69.6 $\pm$ 0.45 & 85.0 $\pm$ 0.80 & 83.0 $\pm$ 1.27 & \textbf{87.1} $\pm$ \textbf{1.32}  \\%& \textbf{67.3} \\
68.6 $\pm$ 0.53 & \textbf{85.8} $\pm$ \textbf{0.62} & 85.7 $\pm$ 2.29 & 85.9 $\pm$ 2.09 \\
64.0 $\pm$ 0.38 & 83.2 $\pm$ 0.20 & \textbf{85.8} $\pm$ \textbf{0.84} & 80.8 $\pm$ 0.93 \\
\bottomrule

\end{tabular} 
\end{sc}
}

\end{small}
\end{center}
%\vspace{-0.4cm}
\end{table}




% %\clearpage

\end{document}